\documentclass[aps,prl,twocolumn,showpacs,preprintnumbers,amsmath,amssymb,superscriptaddress]{revtex4}%

\usepackage{graphicx}%
\usepackage{dcolumn}
\usepackage{amsmath}
\usepackage{color}
\usepackage{bm}

\begin{document}


\title{Superconductivity and Field-Induced Magnetism in SrFe$_{1.75}$Co$_{0.25}$As$_2$}

\author{R.~Khasanov}
 \email[Corresponding author: ]{rustem.khasanov@psi.ch}
 \affiliation{Laboratory for Muon Spin Spectroscopy, Paul Scherrer
Institut, CH-5232 Villigen PSI, Switzerland}
\author{A.~Maisuradze}
 \affiliation{Laboratory for Muon Spin Spectroscopy, Paul Scherrer
Institut, CH-5232 Villigen PSI, Switzerland}
\author{H. Maeter }
 \affiliation{IFP, Technische Universit\"at Dresden, 01069 Dresden, Germany}
\author{A.~Kwadrin}
 \affiliation{IFP, Technische Universit\"at Dresden, 01069 Dresden, Germany}
\author{H.~Luetkens}
 \affiliation{Laboratory for Muon Spin Spectroscopy, Paul Scherrer
Institut, CH-5232 Villigen PSI, Switzerland}
\author{A.~Amato}
 \affiliation{Laboratory for Muon Spin Spectroscopy, Paul Scherrer
Institut, CH-5232 Villigen PSI, Switzerland}
\author{W.~Schnelle}
\affiliation{Max-Planck-Institut f\"ur Chemische Physik fester Stoffe, N\"othnitzer Str.
40, 01187 Dresden, Germany}
\author{H.~Rosner}
\affiliation{Max-Planck-Institut f\"ur Chemische Physik fester Stoffe, N\"othnitzer Str.
40, 01187 Dresden, Germany}
\author{A.~Leithe-Jasper}
\affiliation{Max-Planck-Institut f\"ur Chemische Physik fester Stoffe, N\"othnitzer Str.
40, 01187 Dresden, Germany}
\author{H.-H.~Klauss}
 \affiliation{IFP, Technische Universit\"at Dresden, 01069 Dresden, Germany}

\begin{abstract}
Using muon-spin rotation, we studied  the in-plane
($\lambda_{ab}$) and the out of plane ($\lambda_c$) magnetic field
penetration depth in SrFe$_{1.75}$Co$_{0.25}$As$_2$ ($T_{\rm
c}\simeq13.3$~K). Both $\lambda_{ab}(T)$ and $\lambda_c(T)$ are
consistent with the presence of two superconducting gaps with the
gap to $T_{\rm c}$ ratios $2\Delta/k_BT_{\rm c}=7.2$ and 2.7. The
penetration depth anisotropy
$\gamma_\lambda=\lambda_c/\lambda_{ab}$ increases from
$\gamma_\lambda\simeq2.1$ at $T_{\rm c}$ to 2.7 at 1.6~K. The mean
internal field in the superconducting state increases with
decreasing temperature, just opposite to the diamagnetic response
seen in magnetization experiments. This unusual behavior suggests
that the external field induces a magnetic order which is
maintained throughout the whole sample volume.
\end{abstract}

\maketitle


The discovery of Fe-based high-temperature superconductors (HTS) with critical temperatures above 50 K has triggered a surge of theoretical and experimental studies. Similar to the cuprates, the Fe-based HTS are characterized by a layered structure and by superconducting states which appear upon doping from parent compounds exhibiting long-range antiferromagnetic (AF) order. On the other hand, the metallicity of the parent compounds, as well as the occurrence of superconductivity on a few disconnected pieces on the Fermi surface clearly distinguish them from cuprate HTS. A fundamental question is to understand whether the mechanism(s) leading to the occurrence of the superconducting ground states in both families share common ground. On this frame, one of the most topical issues is to compare the interplay between magnetism and superconductivity between both families. This can be realized by carefully monitoring the emergence of these ground states as a function of chemical doping, applied magnetic field or external pressure.

In lightly doped cuprate HTS, static magnetism is found to coexist with superconductivity on a nanometer scale, as revealed by neutron diffraction, nuclear quadrupolar resonance  and muon-spin rotation ($\mu$SR) experiments \cite{Wakimoto01,Julien99,Niedermayer98,Sanna04}. For this family, stripe-like modulations of the charge and spin densities are found \cite{Sanna04}. On the other hand, for the Fe-based HTS REO$_{1-x}$F$_x$FeAs (RE=La, Ce, etc.) and (Sr,Ba)$_{1-x}$K$_x$Fe$_2$As$_2$ the majority of the reported phase diagram studies show either the occurrence of an abrupt first-order like change, with a full suppression of the AF order, at the doping where superconductivity emerges \cite{Luetkens08_phase-diagramm,Zhao08}, or point to a microscopic separation of magnetism and superconductivity \cite{Rotter08,Goko08}. To date, a microscopic coexistence of both states is solely reported for the system SmFeAsO$_{1-x}$F$_x$ \cite{Drew09,Sanna09}.

The subtle balance between superconductivity and
magnetism in cuprate HTS is strongly affected by the magnetic field.
Field-induced or enhanced static magnetic order are detected in
various underdoped cuprates. In electron doped
Pr$_{2-x}$Ce$_x$CuO$_{4-y}$ \cite{Sonier03} and
Pr$_{1-x}$LaCe$_x$CuO$_{4-y}$ \cite{Kadono0405}, {\it e.g.}, even
a magnetic field  as low as 10~mT is sufficient to enhance the
weak AF ordering over the entire sample volume. For Fe-based
HTS we are aware of only two reports pointing to a
possible enhancement of magnetism by applied magnetic field
\cite{Luetkens08_La1111,Pratt08}. Here, we report on $\mu$SR
studies of a single crystalline sample of
SrFe$_{1.75}$Co$_{0.25}$As$_2$. In zero external field,
superconductivity coexists with dilute Fe spins which are static
on the $\mu$SR time scale. In the superconducting state, the
applied field leads to appearance of an additional source of local
magnetic field, maintained throughout the whole volume of the
sample, thus pointing to the field-induced ordering of the Fe
moments.

\begin{figure}[htb]
\includegraphics[width=0.75\linewidth]{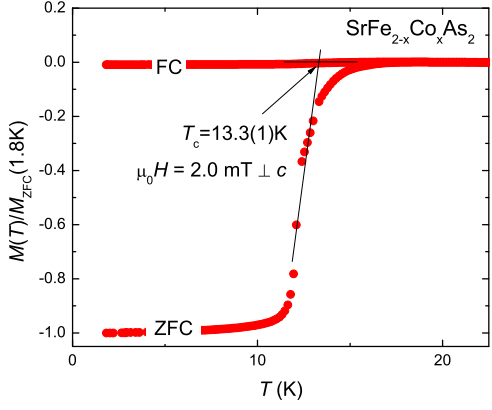}
 \vspace{-0.3cm}
\caption{(Color online) ZFC and FC
magnetization of SrFe$_{1.75}$Co$_{0.25}$As$_2$ at $\mu_0H =
2.0$~mT$\perp c$ (the small paramagnetic contribution is subtracted). $T_c$ is obtained as the intersect of linearly extrapolated
$M_{\rm ZFC}(T)$  with $M=0$ line.}
 \label{fig:magnetization}
\end{figure}

The SrFe$_{1.75}$Co$_{0.25}$As$_2$ (SFCA) single crystalline
samples  were synthesized as described in \cite{Kasinathan09}. The
zero-field cooled (ZFC) and field cooled (FC) magnetization ($M$)
measurements confirm bulk superconductivity (see
Fig.~\ref{fig:magnetization} and Ref.~\onlinecite{Kasinathan09}). The transition temperature ($T_{\rm
c}$) of 13.3(1)~K is consistent with the overdoped composition  of the
sample \cite{Leithe-Jasper08}. The true FC Meissner effect is very
small, suggesting that pinning in SFCA is relatively
strong. The ZFC shielding is $\sim100$\%. All crystals used in our study were taken from the same growth batch and show very similar $M(T)$ dependences.

The $\mu$SR experiments were carried  out at the $\pi$M3 beam line
(Paul Scherrer Institute, Villigen, Switzerland). Three SFCA
single crystals with an approximate size of $2.5\times$2.0$\times$0.1~mm$^3$ were mounted on a holder
designed to conduct $\mu$SR measurements on thin
single-crystalline samples. The zero-field (ZF) and
transverse-field (TF) $\mu$SR experiments were performed at
temperatures ranging from 1.5 to 150~K. In two sets of TF
measurements the external magnetic field $\mu_0H=10$~mT was
applied parallel and perpendicular to the crystallographic
$c$ axis, and always perpendicular to the muon-spin polarization.
The typical counting statistics were $\sim1.5\cdot 10^{7}$
positron events for each particular data point.

\begin{figure}[htb]
\includegraphics[width=0.88\linewidth]{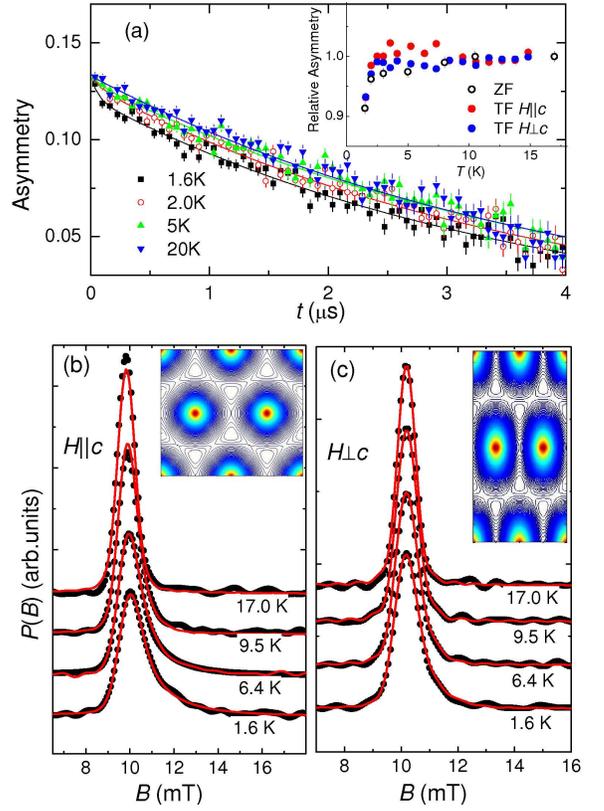}
 \vspace{-0.3cm}
\caption{(Color online) (a) ZF $\mu$SR time-spectra  of
SrFe$_{1.75}$Co$_{0.25}$As$_2$. The lines are fits by
Eq.~(\ref{eq:ZF}). The inset shows the temperature dependences of
normalized asymmetries (see text for details). (b, c) The magnetic
field  distributions $P(B)$ obtained in $H\parallel c$ and $H\perp
c$ set of TF $\mu$SR experiments. The lines are the fits by
Eq.~(\ref{eq:Pt}). The insets represent the contour plots of field
variation.}
 \label{fig:ZF_TF}
\end{figure}

In ZF, the muon-spin polarization is relaxed by magnetic moments
of electronic and nuclear origin. As shown in
Fig.~\ref{fig:ZF_TF}a, the relaxation rate of the ZF-$\mu$SR
signal is constant down to $T\simeq5$~K, while at lower temperatures an
additional fast relaxing component starts to develop. The solid
lines in Fig.~\ref{fig:ZF_TF}a correspond to the fit by:
\begin{equation}
A^{\rm ZF}(t)=A_1 \exp(-\Lambda_1 t)+A_2 \exp(-\Lambda_2 t)
 \label{eq:ZF}
\end{equation}
Here $A_1$ and $A_2$ are the initial asymmetries of the slow  and
the fast relaxing components, and $\Lambda_1$ and $\Lambda_2$ are
the corresponding exponential depolarization rates. The
temperature dependence of $A_1$ normalized to the total ZF
asymmetry $A_1+A_2$ is shown in the inset. The fit reveals that
$\Lambda_1$ is temperature independent which is also clearly seen
from the raw data as a parallel shift of $A^{\rm ZF}(t)$ at
$t\gtrsim0.2$~$\mu$s with decreasing temperature. Measurements in
a longitudinal-field geometry indicate that the exponential
character of the muon-spin relaxation is due to randomly oriented
local magnetic fields, which are static on the $\mu$SR time scale.
Such behavior is consistent with dilute Fe moments as observed
recently for another representative of Fe-based HTS
FeSe$_{0.85}$ \cite{Khasanov08_FeSe}.

In the TF geometry  muons measure the magnetic field distribution
[$P(B)$] inside the sample. For the superconductor in the vortex
state, $P(B)$ is uniquely determined by the magnetic penetration
depth $\lambda$ and the coherence length $\xi$ \cite{Yaouanc97}.
Few representative $P(B)$ distributions, obtained after Fourier
transformation of TF $\mu$SR time-spectra, are presented in
Figs.~\ref{fig:ZF_TF}b, c.
In the normal state, a symmetric line at the position of the
external magnetic field with a broadening arising from the nuclear
and electronic magnetic moments is seen. Below $T_{\rm c}$, the
field distribution is broadened and asymmetric which is
characteristics of the inhomogeneous field distribution within the
flux line lattice (FLL).

In an orthogonal reference frame $x,y,z$ with $H\parallel z$  ($z$
is one of the principal axes $a$, $b$, or $c$) the spatial
magnetic field distribution within a FLL of an anisotropic
superconductor is \cite{Yaouanc97}:
\begin{equation}
B({\bf{r}}) = \langle B \rangle \sum_{\bf{G}} \exp(-i{\bf{G \cdot
r}})  B_{\bf{G}}(\lambda_{x}, \lambda_y, \xi_x, \xi_y).
 \label{eq:Field-distribution}
\end{equation}
{\bf G} is the reciprocal lattice vector, $\langle B \rangle$ is
the average magnetic field inside the superconductor, $\bf{r}$ is
the vector coordinate in a plane perpendicular to the applied
field, and the Fourier components $B_{\bf{G}}$, obtained within the
framework of the Ginzburg-Landau (GL) model, are \cite{Yaouanc97}:
\begin{equation}
B_{\bf{G}} = \frac{\Phi_0}{S} (1-b^4)\frac{uK_1(u)}{\lambda_x^2G_y^2 + \lambda_y^2G_x^2}.
 \label{eq:Field-distribution2}
\end{equation}
Here, $\Phi_0$ is the magnetic flux quantum, $S=\Phi_0/\langle
B\rangle$ is the FLL unit cell area, $b = \langle B\rangle/B_{\rm
c2}$ ($B_{\rm c2}$ is the second critical field), $K_1(u)$ is the
modified Bessel function, and $u^2 = 2(\xi_x^2G_x^2 +
\xi_y^2G_y^2)(1+b^4)[1-2b(1-b)^2]$. The reciprocal lattice
corresponding to the field distribution in hexagonal FLL is: ${\bf G}_{m, n} =
(2\pi /S)[ym,  (n - m/2)x]$, where $x = (2S\lambda_x/\sqrt{3} \lambda_y)^{1/2}$, $y=(\sqrt{3}S\lambda_y/2 \lambda_x)^{1/2}$, and $m$ and $n$ are the integer
numbers \cite{Thiemann89}. Note that in the  uniaxial case
($\lambda_x=\lambda_y$), Eq.~(\ref{eq:Field-distribution2})
converts into the standard GL equation for an isotropic
superconductor \cite{Yaouanc97}.

The TF $\mu$SR time spectra were fitted to a theoretical asymmetry
function $A^{\rm TF}(t)$ by assuming the internal field
distribution $P_{\rm id}(B)$ obtained from
Eq.~(\ref{eq:Field-distribution}) and accounting for the FLL
disorder and the electronic moment contributions (see the ZF
discussion above) by convoluting $P_{\rm id}(B)$ with Gaussian
and Lorenzian functions:
\begin{equation}
A^{\rm TF}(t) = Ae^{i\phi} e^{-\sigma_g^2t^2/2-\Lambda t} \int P_{\rm
id}(B)e^{i \gamma_{\mu}Bt}dB.
 \label{eq:Pt}
\end{equation}
Here $A$ and $\phi$ are the initial asymmetry and the phase of the
muon spin ensemble, $\sigma_g$ accounts for the  FLL disorder
\cite{Maisuradze08}, and $\Lambda$ relates to the electronic moment
contribution and is assumed to be  equal to $\Lambda_1$ obtained
in ZF experiments. During the analysis we first fit the data
measured in $H\parallel c$ orientation which, considering
$\lambda_a=\lambda_b$, allows to obtain the in-plane magnetic
penetration depth $\lambda_{ab}$. Note that at such a low field as
$\mu_0H=10$~mT, $P(B)$ is independent on $\xi$
\cite{Maisuradze08}. $\lambda_{ab}$ was further used in the fit of
$H\perp c$ set of data by assuming in
Eq.~(\ref{eq:Field-distribution2}) $\lambda_x=\lambda_{ab}$ and
$\lambda_y=\lambda_{c}$. The resulting fitted curves are
represented in Figs.~\ref{fig:ZF_TF}b and c by red lines.
The field distributions obtained by
Eq.~(\ref{eq:Field-distribution}) are shown in the insets.

The temperature dependences of the initial asymmetries are shown
in the inset of Fig.~\ref{fig:ZF_TF}a. Both, $A^{\parallel c}(T)$
and $A^{\perp c}(T)$ are almost constant down to $T\simeq 3$~K and
decrease by $\sim10$~\% at $T=1.6$~K just resembling the
temperature behavior of the slow relaxing component $A_1$ observed
in ZF measurements. This suggests that the fast relaxing
component, seen in ZF $\mu$SR experiments, appears from areas
which are separated in space from the ``slow-relaxing'' ones.  The
superconductivity in such magnetic regions may either coexist with
magnetism on a nanometer scale \cite{Sanna09}, or become suppressed due
to enhanced magnetic order
\cite{Goko08,Khasanov09-anisotropy}.
\begin{figure}[htb]
\includegraphics[width=0.88\linewidth]{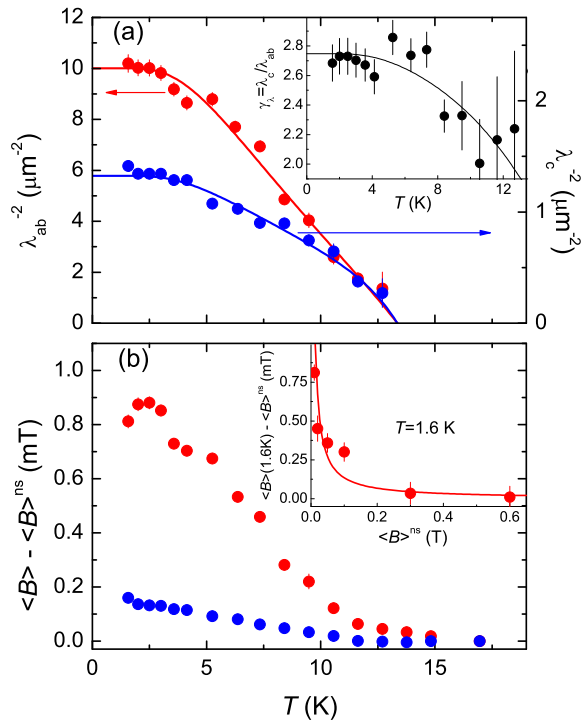}
 \vspace{-0.3cm}
\caption{(Color online) (a) Dependence of $\lambda_{ab}^{-2}$  and
$\lambda_c^{-2}$ on temperature of SrFe$_{1.75}$Co$_{0.25}$As$_2$.
The solid lines are the fits by Eq.~(\ref{eq:lambda_ab}) with the
parameters summarized in Table~\ref{Table:lambda_results}. The
inset shows the temperature dependence of the penetration depth
anisotropy $\gamma_\lambda=\lambda_c/\lambda_{ab}$. (b)
Temperature dependence of $\langle B\rangle-\langle B\rangle^{\rm
ns}$ obtained in $H\parallel c$ (red symbols) and $H\perp c$ (blue
symbols) set of experiments. The inset shows the dependence of
$\langle B\rangle (1.6$~K$)-\langle B\rangle^{\rm ns}$ on $\langle
B\rangle^{\rm ns}$ in $H\parallel c$ orientation.}
 \label{fig:lambda(T)_B(T)}
\end{figure}

The dependence of $\lambda^{-2}_{ab}$ and $\lambda^{-2}_{c}$ on
temperature is shown in Fig.~\ref{fig:lambda(T)_B(T)}a. The experimental
data were analyzed within the framework of
the phenomenological $\alpha$-model by assuming two independent
contributions to $\lambda^{-2}$ \cite{Carrington03}:
\begin{equation}
\frac{\lambda^{-2}(T)}{\lambda^{-2}(0)}=
\omega\cdot\frac{\lambda^{-2}(T,
\Delta_{0,1})}{\lambda^{-2}(0,\Delta_{0,1})}+(1-\omega)\cdot
\frac{\lambda^{-2}(T, \Delta_{0,2})}{\lambda^{-2}(0,\Delta_{0,2})}
 \label{eq:lambda_ab}
\end{equation}
In this model, each superconducting gap, $\Delta_1(T)$ and
$\Delta_2(T)$, has a similar temperature dependence given by
$\Delta_i(T)=\Delta_{i,0}\tanh\{1.82[1.018(T_{\rm
c}/T-1)]^{0.51}\}$, Ref.~\onlinecite{Carrington03}, but different
zero-temperature values ($\Delta_{0,i}$, $i=1$, 2). In our
analysis both the large and the small gap were assumed to be of
$s$-wave symmetry. The parameter $\omega$ accounts for the
relative contribution of the larger gap to $\lambda^{-2}$ and
$\lambda^{-2}(0)$ is the penetration depth at $T=0$. Each
component $\lambda^{-2}(T,
\Delta_{0,i})/\lambda^{-2}(0,\Delta_{0,i})$ is calculated within
the local (London) approximation \cite{Khasanov08_Inf-Layer}.
During the fit, both $\lambda_{ab}(T)$ and $\lambda_c(T)$ were
assumed to be described by the same small and large gaps
($\Delta_{1,ab}=\Delta_{1,c}$ and $\Delta_{2,ab}=\Delta_{2,c}$),
but different weighting factors ($\omega_{ab}\neq\omega_c$). The
results of the fit are summarized in Fig.~\ref{fig:lambda(T)_B(T)}a and
Table~\ref{Table:lambda_results}. The temperature dependence of
the  anisotropy parameter $\gamma_\lambda=\lambda_c/\lambda_{ab}$
is shown in the inset. Four important results are deduced from this analysis:
(i) The contribution of the large gap to $\lambda_{ab}^{-2}(T)$ is
relatively small in comparison with that observed in
Ba$_{1-x}$K$_x$Fe$_2$As$_2$ \cite{Khasanov09-anisotropy}. This may
be explained by the fact that the electron doping due to Co
substitution reduces the size of the Fermi surface pockets at the
Brillouin zone center (where the large gap opens), but leads to substantial
enhancement of the pockets at the zone corner (where the small gap develops)
\cite{Sekiba08,Terashima08}.
(ii) $\gamma_\lambda$ increases from
$\gamma_\lambda\simeq2.1$ at $T_{\rm c}$ to 2.7 at 1.6~K. This
increase is consistent with the general trend obtained for various
Fe-based HTS
\cite{Weyeneth08_Tanatar08,Khasanov09-anisotropy}. Our data
imply that the different
temperature dependences of $\lambda^{-2}_{ab}$ and
$\lambda_{c}^{-2}$, as well as dependence of $\gamma_\lambda$ on
$T$, are due to much smaller contribution of the larger gap to
$\lambda_{ab}^{-2}$ than that to $\lambda_{c}^{-2}$.
(iii) $\gamma_\lambda$ is very close to the calculated ratio of the plasma frequencies
$\gamma_{\omega_p}=\omega_p^a/\omega_p^c\simeq 2.8$ of Sr$_2$Fe$_2$As$_2$ \cite{Kasinathan09}. Note that within the London theory $\gamma_\lambda\equiv\gamma_{\omega_p}$.
(iv) The ratios $2\Delta_{0,1}/k_BT_c=7.2$ and
$2\Delta_{0,2}/k_BT_{\rm c}=2.7$ are well within the ranges
established for various Fe-based HTS. As shown in
Ref.~\cite{Evtushinsky08_gaps}, Fe-based HTS  bear two
nearly isotropic gaps with $2\Delta/k_BT_{\rm c}=7\pm 2$ and
$2.5\pm 1.5$.
\begin{table}[htb]
\caption[~]{\label{Table:lambda_results} Summary of
$\lambda_{ab}^{-2}(T)$ and $\lambda_c(T)^{-2}$ study of
SrFe$_{1.75}$Co$_{0.25}$As$_2$ (see text for details).
} %
\begin{center}
\vspace{-0.2cm}
\begin{tabular}{ccccccccc}
\hline
\hline
 &$T_{\rm c}$ &$\Delta_{0,1}$&$\frac{2\Delta_{0,1}}{k_BT_{\rm c}}$&$\Delta_{0,2}$&$\frac{2\Delta_{0,2}}{k_BT_{\rm c}}$&$\omega$&$\lambda(0)$\\
&(K)&(meV)&&(meV)&&&(nm)\\
\hline

$\lambda_{ab}^{-2}(T)$ &13.35$^{\rm a}$& 4.14$^{\rm a}$&7.2 &1.56$^{\rm a}$& 2.7 &0.04 & 315\\
$\lambda_{c}^{-2}(T)$  &13.35$^{\rm a}$& 4.14$^{\rm a}$&7.2 &1.56$^{\rm a}$& 2.7 &0.29 & 870\\

 \hline \hline

\end{tabular}
   \end{center}
   \vspace{-0.3cm}
   $^a${\small Common parameters used in the analysis}
\end{table}

In Figure~\ref{fig:lambda(T)_B(T)}b we plot the difference between
the internal magnetic field $\langle B\rangle$ and that measured
in the normal state at $T\simeq20$~K ($\langle B\rangle^{\rm
ns}$). In contrast to what is expected for a superconductor,
$\langle B\rangle$ {\it increases} with decreasing temperature. We
may clearly rule out the possibility to explain the positive field
shift by a paramagnetic Meissner effect, since both ZFC and FC
magnetization result in a diamagnetic shift (see
Fig.~\ref{fig:magnetization}). Also, it cannot be explained by
the reduction of a hypothetical negative muon Knight shift due to
condensation of the carriers into Cooper pairs
\cite{Feyerherm94}, otherwise the difference between $\langle
B\rangle$(1.6~K), measured deeply in the superconducting state,
and $\langle B\rangle^{\rm ns}$ would increase with increasing
field. The inset in Fig.~\ref{fig:lambda(T)_B(T)}b shows, in
contrast, that $\langle B\rangle (1.6$~K$)-\langle B\rangle^{\rm
ns}$ decreases.

Both, the temperature and the magnetic  field dependence of
$\langle B\rangle-\langle B\rangle^{\rm ns}$ resemble the
situation in the electron doped cuprate HTS
Pr$_{2-x}$Ce$_x$CuO$_{4-y}$ \cite{Sonier03} and
Pr$_{1-x}$LaCe$_x$CuO$_{4-y}$ \cite{Kadono0405}. Those reports are
different in some details, but agreed that the paramagnetic
response in the superconducting state, seen by muons, is caused by
the field-induced weak AF order. The external field
applied in $c$ direction leads to the appearance of an additional
field component at the muon stopping site, which is {\it
perpendicular} to the $c$ axis
\cite{Sonier03,Kadono0405}.
We may assume that the paramagnetic response  of SFCA, seen in
Fig.~\ref{fig:lambda(T)_B(T)}b, is caused by the same mechanism.
The solid line in the inset of Fig.~\ref{fig:lambda(T)_B(T)}b
corresponds to the fit by $\langle B\rangle (1.6$~K$)=[(\langle
B\rangle^{\rm ns})^2+(B^\perp)^2]^{1/2}$ \cite{Sonier03} with the
induced field component $B^\perp=5.2$~mT. We emphasize,
that the field-induced AF order in SFCA is different from that
for the parent compound SrFe$_2$As$_2$, where the
magnetic field on the muon site is {\it parallel} to the $c$ axis
\cite{Luetkens09_muon-site}.

The strong influence of the superconducting phase is observed at
external field as low as 10~mT, which corresponds to the
intervortex distance $\sim 500$~nm. Considering that most of the
implanted muons probe the sample outside of the vortex cores, we
may conclude that the field-induced AF order extends throughout
the whole sample volume.
Further experimental and theoretical studies  are needed in order
to elucidate the origin of these effects, as well as to consider a
possible role of muons as a source of perturbation.


To conclude, muon-spin rotation measurements  were performed on a
single-crystalline sample of SrFe$_{1.75}$Co$_{0.25}$As$_2$ ($T_{\rm
c}\simeq13.3$~K). In zero field,  superconductivity
coexists with dilute Fe moments which are static on the $\mu$SR
time scale.
The temperature dependences of the in-plane $\lambda_{ab}$ and the
out of plane $\lambda_c$ magnetic penetration depth are well
described assuming the presence of two $s-$wave like gaps
with zero-temperature values $\Delta_{0,1}=4.14$~meV and
$\Delta_{0,2}=1.56$~meV, respectively. The gap to $T_{\rm c}$
ratios are $2\Delta/k_BT_{\rm c}=7.2$ and 2.7, in agreement with
those reported for various Fe-based HTS
\cite{Evtushinsky08_gaps}.
The mean internal field in the superconducting state increases
with decreasing temperature, just opposite to the diamagnetic
response seen in magnetization experiments. This may be due to
weak field-induced antiferromagnetic order leading to appearance
of in-plane magnetic field component on the muon stoping site.
The fact that the magnetism is induced by a field as low as 10~mT
points to strong interplay between the magnetic and the
superconducting order parameters in Fe-based HTS.
Whether these order parameters coexist or separate on a
microscopic level is still an open question.


The work was performed at the Swiss Muon Source (S$\mu$S),  Paul
Scherrer Institute (PSI, Switzerland).

\end{document}